# Similarities between the generation and dynamics of concentric and non-concentric multiple double layers

G. Amarandei[a,b], C. Gaman[a,b], D. G. Dimitriu[b], C. Ionita[a],

M. Sanduloviciu[b], R. Schrittwieser[a]

[a]*Institute for Ion Physics, Leopold-Franzens University, 25 Technikerstr., A-6020 Innsbruck, Austria, e-mail:* Roman.Schrittwieser@uibk.ac.at

[b]*Faculty of Physics, "Al. I. Cuza" University, 11 Carol I Blvd., RO-700506 Iasi, Romania, e-mail:* dimitriu@uaic.ro

*Abstract*

We report new experimental results on the generation and dynamics of multiple double layers in front of a positively biased electrode immersed in a DP-machine plasma. Depending on the geometry or the dimensions of the electrode, this complex space charge configuration can appear in two different modes. If the electrode is small with respect to the plasma size, the phenomenon appears as a set of concentric luminous shells with different intensities. If the electrode has large dimensions, or a one-dimensional geometry, the spherical double layers appear adjacent to each other, as equally separated luminous plasma blobs.

Our experimental investigations prove that in both of cases we deal with almost the same phenomenon. The same elementary processes, namely electron-neutral impact excitations and ionizations, play the key role in the generation and dynamics of each structure. This is emphasized in the static current-voltage characteristic of the electrode, which shows similar jumps of the current, associated with hysteresis.

## Introduction

Double layers (DLs) are nonlinear potential structures consisting of two adjacent layers of positive and negative space charge, respectively. Between these layers a potential jump exists, creating an electric field. A common way to obtain a DL structure is to positively bias an elec-trode immersed in a plasma being in a steady state. In this case, a complex space charge configuration (CSCC) in form of a quasi-spherical luminous plasma body attached to the electrode is obtained [1-3]. Experimental investigations have revealed that such a CSCC consists of a positive "nucleus" (an ion-rich plasma) bordered by a nearly spherical electrical DL [2-6]. The potential drop across the DL is almost equal to the ionization potential of the background gas atoms.

Under certain experimental conditions (plasma density, gas nature and pressure, electron temperature, geometry of the experiment) a more complex structure in form of two or more subsequent DLs was observed [7-16]. It appears as several bright and concentric plasma shells attached to the anode of a glow discharge or to a positively biased electrode immersed in plasma. The successive DLs are precisely located at the abrupt changes of luminosity between two adjacent plasma shells. Emissive probe measurements have shown that the axial profile of the plasma potential has a stair step shape, with potential jumps close to the ionization potential of the used gas [12,16]. The static current-voltage characteristic of the electrode has demonstrated that each of the DLs appears simultaneously with a current jump [16].

However, if the electrode is large or asymmetric (with one dimension larger than the other one), a multiple double layer (MDL) structure appears non-concentrically, as a network of plasma spots, almost equally distributed on the electrode surface [17-19]. Each of the plasma spots is a CSCC as described above. Here, we will present experimental results which prove that concentric and non-concentric MDLs are similar structures, the same elementary processes being implicated.

**Experimental results and discussion**

The experiments were performed in the DP (double plasma) machine of the University of Innsbruck, extensively described in [16]. We used only the target chamber of the machine. The plasma was pulled away from its steady state by gradually increasing the voltage of a rectangular tantalum electrode, 6 cm long and 0.5 cm width. The background argon pressure was $p \cong 5\times 10^{-3}$ mbar and the plasma density $n \cong 10^{10}$-$10^{11}$ cm$^{-3}$. An XY recorder was used to register the static current-voltage characteristic of the electrode. The ac component of the electrode current was recorded by using a digital computer-controlled oscilloscope. The plasma potential was measured by an emissive probe. Fig. 1 shows the static current-voltage ($I - V$) characteristic of the electrode, obtained by gradually increasing and subsequently decreasing the voltage of the power supply, for the case where in front of the electrode two plasma spots appear. Thus, by increasing the potential on the electrode, at a critical value of it (corresponding to the point **a** on the static $I – V$ characteristic from the Fig. 1) at first a luminous spot appears on a certain point of the electrode (Fig. 2a), where the current is highest because of local causes (for example the presence of protuberances on the electrode surface or local gas emission). By further increasing the potential on the electrode, a second luminous spot appear on the electrode (Fig. 2b) simultaneously with a new jump of the current (**c → d** in Fig. 1). When the voltage is gradually decreasing we observe that both of the current jumps are associated with hysteresis effects (**a → b → g → h** and **c → d → e → f**, respectively in Fig. 1). This proves the capability of each of the plasma spots to maintain by itself under conditions less favorable than those required for their appearance. Emissive probe measurements revealed that each plasma spot consist of a positive "nucleus" (an ion-rich plasma) surrounded by a DL.

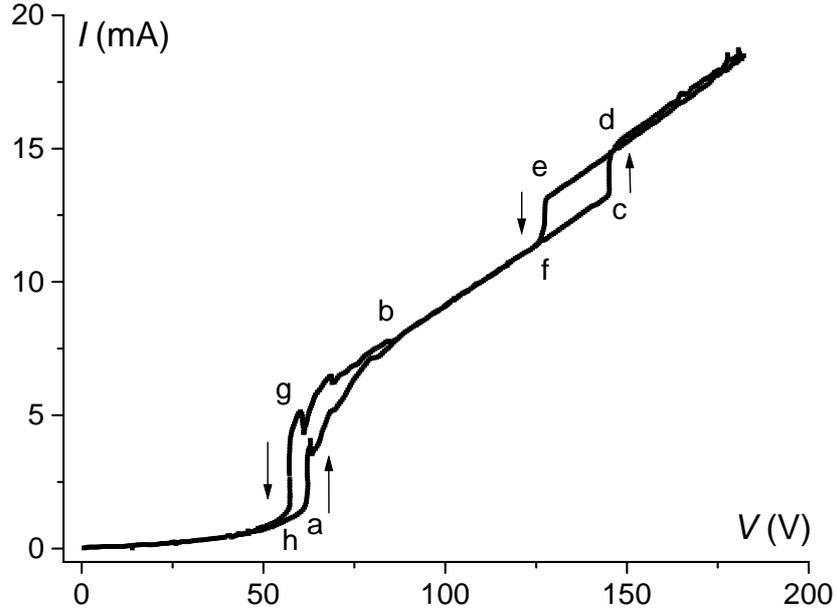

Fig. 1: Static current-voltage characteristic of the electrode obtained for the conditions where two plasma spots appear.

The obtained experimental results show many similarities with those obtained in the case of the generation of a concentric MDL [16]. Thus, in both of cases each of the DL structures appears simultaneously with a jump of the current collected by the electrode, associated with hysteresis effect. This proves that the same elementary processes are involved in the MDL's generation, namely electron-neutral impact excitations and ionizations. These processes determine the appearance of negative and positive space charges, respectively, spatially well separated because of the dependence of the respective cross sections on the kinetic electron energies, which varies along the structure according to the potential profile. When the po-tential drop between the opposite space charge attains the ionization potential of the gas, the DL structure is formed. Further, depending on the geometry of the experiment, a concentric or non-concentric MDL will form. In the case of a non-concentric MDL, the plasma spots are almost equally distributed on the electrode surface because of the electrostatic forces that appear between their negative external shells. The size and the number of plasma spots (in the case of non-concentric MDLs), as well as the size and the number of plasma shells (in the case of concentric MDLs) depend on the electrode potential and the discharge current in the same way. Thus, a larger potential on the electrode increases the number of DLs, while an increase of the discharge current plays the opposite role.

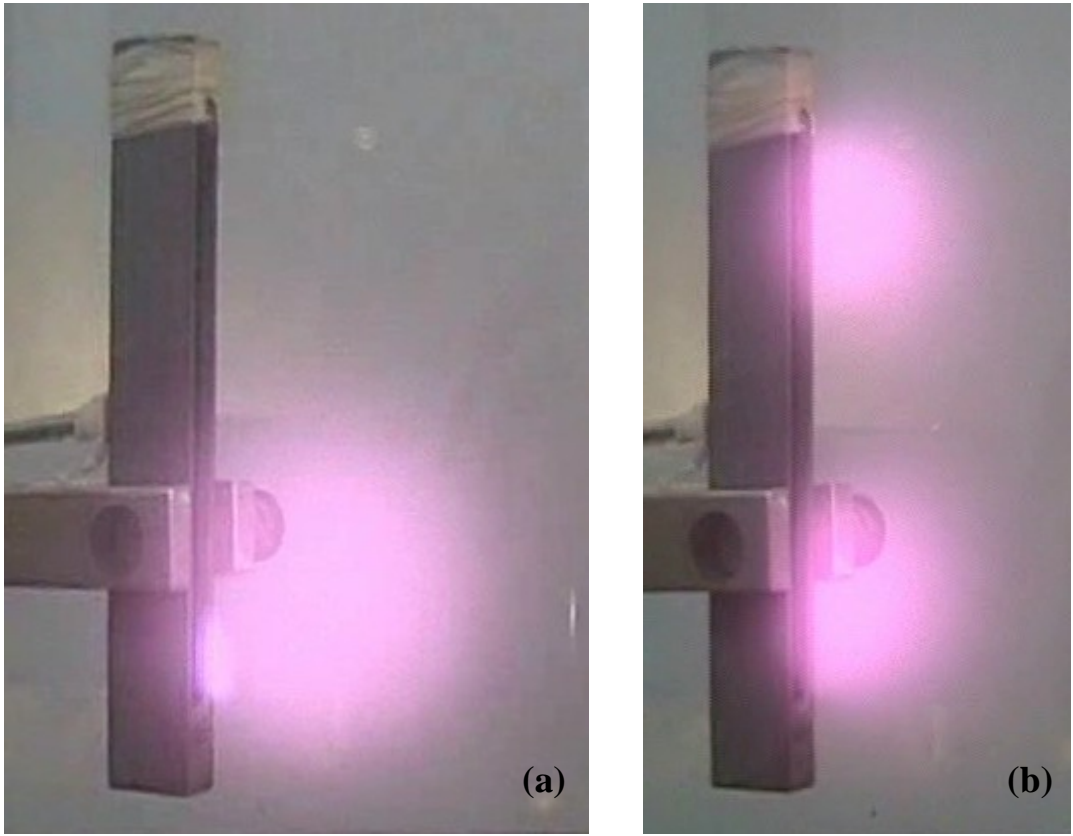

Fig. 2: Luminous plasma spots formed on a positively biased asymmetric (rectangular) electrode immersed into a plasma.

## Conclusion

Experimental results prove that a common mechanism is active at the origin of the emergence of concentric and non-concentric multiple double layers in plasmas. The important role of the electron-neutral impact excitations and ionizations for the formation of such complex structures was demonstrated.